\documentclass[aps,prd,twocolumn,groupedaddress]{revtex4}
\usepackage{amsmath, amsfonts, graphicx}

\begin{document}

\title{The Fourth Generation t-prime in
Extensions of the Standard Model}

\author{Erin De Pree}\email[]{ekdepree@smcm.edu}
\affiliation{Department of Physics, St. Mary's College of Maryland, St. Mary's City, MD 20686, USA}

\author{Gardner Marshall}\email[]{grmarshall@wm.edu}
\author{Marc Sher}\email[]{mtsher@wm.edu}
\affiliation{Particle Theory Group, Department of Physics, College of William and Mary, Williamsburg, VA 23187, USA}

\date{\today}

\begin{abstract}
    We study the effects of a fourth generation $t^{\prime}$ quark in various extensions of the standard model.  In the
    Randall-Sundrum model, the decay $t^{\prime} \rightarrow t Z$ has a large branching ratio that could be detected
    at the Large Hadron Collider (LHC).   We also look at the two-Higgs doublet models I, II and III, and note that, in
    the latter, the branching ratio of $t^{\prime} \rightarrow t \phi$, where $\phi$ is a Higgs scalar or pseudoscalar, is
    huge and we discuss detection at the LHC.    A few comments about the minimal supersymmetric standard
    model are also included.
\end{abstract}



\maketitle


\section{Introduction}

    Interest in a sequential fourth generation has waxed and waned over
    the years \cite{fhs,holdom}. Shortly after the discovery of the third
    generation, a fourth generation was an obvious extension.  However,
    interest in a fourth generation dropped substantially after measurement of the number
    of light neutrinos at the $Z$ pole showed that only three light neutrinos
    could exist.   The discovery of neutrino oscillations suggested the
    possibility of a mass scale beyond the standard model, and models with
    a fourth generation containing a sufficiently massive neutrino became
    acceptable.   In the early part of this decade, it was thought \cite{pdg}
    that electroweak precision measurements ruled out a fourth generation,
    however it was subsequently pointed out \cite{subsequent} that if
    the fourth generation masses are not degenerate, then these
    constraints can be evaded.  More recently, Kribs et al. \cite{kpst}
    showed that a mass splitting of $40-60$ GeV between the fourth
    generation quarks
    results in S and T parameters which are within the one-sigma error
    ellipse.

    Most analyses of the phenomenology of a sequential fourth generation
    have focused on the minimal Standard Model.   In this paper, we
    consider the phenomenology of the fourth generation $t^{\prime}$ in
    popular extensions of the Standard Model.   In Section II, we discuss the
    Randall-Sundrum model and show that one expects a relatively large
    branching ratio for the flavor-changing decay
    $t^{\prime}\rightarrow t Z$, which could be detected at the LHC.  In
    Section III, we study the two-Higgs doublet models (Models I, II and
    III), and show that in Model III, the decay $t^{\prime}\rightarrow t
    \phi$, where $\phi$ is either a Higgs scalar or pseudoscalar, can
    have a huge branching ratio (as high as 95\%)  and we discuss the
    rather dramatic phenomenology at the LHC.  Finally, in Section IV,
    we present our conclusions.

\section{Randall-Sundrum Model}

    The Randall-Sundrum model (RS1) \cite{Randall} is a popular solution
    to the hierarchy problem, where the warped geometry of an
    additional dimension is responsible for generating TeV scale physics
    from a more fundamental Planck scale. 
    In the original RS1 model, the standard model (SM) fields were
    confined to the TeV brane, but it was quickly noted that they could
    be placed in the bulk without problems \cite{Gherghetta,
    Huber:2003tu, Huber:2000ie}. This led
    to a natural resolution of the flavor hierarchy problem. 

    In the basis of diagonal bulk masses, the normalized wavefunction of
    zero-mode fermions is given by \cite{Gherghetta,
    Huber:2003tu, Huber:2000ie}
    \begin{equation}
        \label{fermion wavefunction}
        f^{(0)}(c_{f},z) = \left[\frac{k (1 - 2c_{f}) (kz)^{(1 - 2c_{f})}}{e^{\beta(1 - 2c_{f})} - 1}\right]^{1/2},
    \end{equation}
    where $\beta = kR\pi \approx 37$, $1/k \leq z \leq e^{\beta}/k$ 
    and $c_{f}$ is the 5-D mass parameter describing the location of 
    the fermion in the bulk.  For $c_{f} < 1/2 \ (c_{f} > 1/2)$ the fermion $\psi_{f}$
    lives near the TeV (Planck) brane
    since its resulting Yukawa coupling with the Higgs field becomes
    large (small). Since the Kaluza-Klein (KK) modes lie near the TeV brane, the
    heavier fermions have larger couplings to KK bosons. As noted by
    Agashe et al. \cite{AgasheShort, AgasheLong}, there is mixing between the
    $Z$-boson and the KK-$Z$ boson, resulting in a shift of the
    fermionic couplings to the $Z$.  
    
    In this model, the portion of the Lagrangian responsible for SM
    flavor violation is given by \cite{AgasheLong}
    \begin{equation}
        \label{flavor violating lagrangian}
        \mathcal{L}^{Z} \sim g_{z} \beta \Delta Z^{\mu} \sum_{f} \overline{\psi}^{(0)}_{f}
            \gamma_{\mu} \frac{1}{f^{2}_{f}} (v_{f} - a_{f}\gamma_{5})\psi^{(0)}_{f},
    \end{equation}
    where $g_{z} = \frac{g_{2}}{2\cos\theta_{W}}$ and $\Delta = \left(\frac{M_{Z}}{M_{KK}}\right)^{2}$.
    The vector and axial vector coefficients depend on the fermion $\psi_{f}$ and are given by
    $v_{f} = T^{f}_{3} - Q^{f}_{3}\sin^{2}\theta_{W}$, and $a_{f} = T^{f}_{3}$. The constants $f_{f}$
    are given in terms of equation \ref{fermion wavefunction} by $\sqrt{2k}/f_{f} = f^{(0)}(c_{f},e^{\beta}/k)$.
    Lastly, $\psi^{(0)}_{f}$ is the zero-mode of the 4-D fermion field in the
    basis of diagonal 5-D bulk masses. However, this basis is not the
    same as the basis in which the 4-D mass terms are diagonal. This
    gives rise to flavor violating terms in the 4-D basis, which have
    larger couplings for heavier fermions.

    Using this formalism, Agashe et al. \cite{AgasheShort} point out that with three
    generations, the branching ratio for $t \rightarrow cZ$ is of
    $\mathcal{O}(10^{-5})$ for a KK-Z mass of 3 TeV. This is a
    significant increase from the SM value. In the case of a fourth
    generation however, there are no electroweak precision constraints
    and so very large flavor changing neutral currents (FCNC) are
    allowed. In particular, one expects to observe $t' \rightarrow tZ$
    at a large rate.   Due to the fact that the $t'$ and the $b'$
    have nearly degenerate masses, the decay of the $t'$ is dominated by $t'\rightarrow Wb$
    unless the mixing angle is very small.
    Thus the decay rate for $t' \rightarrow Wb$ is proportional to
    $|V_{t'b}|^{2} \sim |(U_{f})_{34}|^{2}$, where $(U_{f})$ is the
    mixing matrix that arises when expanding out the 4-D fermion fields
    from equation \ref{flavor violating lagrangian} in the basis of
    diagonal 4-D mass terms. Thus the $|V_{t'b}|^{2}$ factor in
    $\Gamma(t' \rightarrow Wb)$ conveniently cancels the
    $|(U_{f})_{34}|^{2}$ factor in $\Gamma(t' \rightarrow tZ)$ when
    calculating the branching ratio BR($t' \rightarrow tZ$).

    The  decay rate and branching ratio for $t' \rightarrow tZ$
    depend on the four 5-D mass parameters for the left and right handed
    $t$ and $t'$.  Two of these can be eliminated in favor of the $t$ and
    $t'$ masses.  Using the central value of $c_{t_{L}}$ given by Agashe
    \cite{AgasheLong}, the number
    of free parameters is reduced to the $t'$ mass and the fermion
    mass parameter $c_{t^{\prime}_{L}}$, which describes the location of
    the $t'$ in the bulk.

    The results for the branching ratio are given in Figure 1 for $t'$
    masses of $400$ and $500$ GeV.   As one varies the third generation
    mass parameter $c_{t_{L}}$
    parameter (discussed in the previous paragraph) over a reasonable
    range, this result changes by less than a factor of two.  We find a
    branching ratio of  $\mathcal{O}(10^{-3}-10^{-2})$.

    ATLAS \cite{atlasfcnc} has claimed that a bound of $10^{-5}$ can be reached in
    $100 \, \textrm{fb}^{-1}$ for $t \rightarrow cZ$.  Since the $Z$ energy in
    the $t'\rightarrow tZ$ decay is similar to that in $t\rightarrow cZ$,
    one can get a rough estimate of the sensitivity by simply scaling this by
    the production cross section.  This gives a sensitivity
    of $10^{-3}$ for $t' \rightarrow tZ$.  One can probably do
    substantially
    better if one includes the fact that the $t$ in the decay can be
    detected, which will help eliminate backgrounds.  
      It is clear that this
    places the decay within reach of the LHC, although a more
    precise analysis would be welcome.

    \begin{figure}
        \includegraphics[scale=.75]{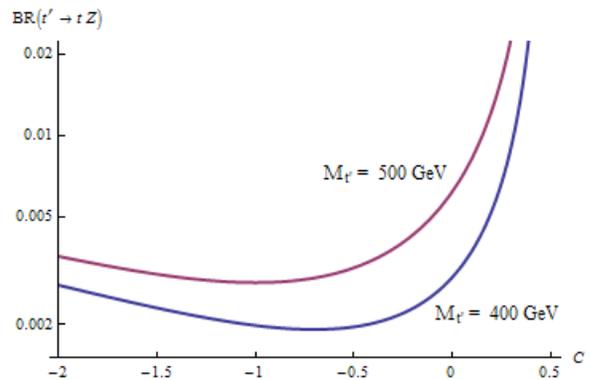}
        \caption{\label{Plot1}Branching Ratio for $t' \rightarrow tZ$ as a function of the
            fermion mass parameter $c_{t^{\prime}_{L}}$.}
    \end{figure}

    How does this branching ratio compare with other models? The most 
    comprehensive study concerning FCNC in the four generation Standard
    Model is the work of Arhrib and Hou \cite{arhribhou}.  They
    considered loop induced FCNC decays of fourth generation quarks,
    plus effects of fourth generation quarks on FCNC decays of third
    generation quarks.  They found that the decay $t' \rightarrow
    tZ$ can occur in the standard model with a branching ratio of
    $\mathcal{O}(10^{-5}-10^{-4})$ and suggest that this may be
    measurable at the LHC.  In the Randall-Sundrum case, the
    branching ratio will be larger by approximately a factor of 100.  
    What about flavor models?  We know of no such models for four 
    generations.  One of us (MS) is studying such a model, but the 
    rate will be similar to that of Arhrib and Hou (additionally, flavor 
    models do not typically give large rates for $t\rightarrow cZ$).  
    It is also likely that other models will have a rate for 
    $t'\rightarrow t\gamma$ or $t'\rightarrow tg$ which is comparable 
    to $t'\rightarrow tZ$, unlike the Randall-Sundrum case.


\section{Two Higgs Doublet Models}

    Among the most popular extensions of the Standard Model are two
    Higgs doublet models \cite{hhg}.   The most common is called Model II.  In the
    two Higgs doublet model II, due to a discrete symmetry,
    the down-type quarks and leptons couple
    to one complex doublet $\phi_{1}$ while the up-type quarks and
    neutrinos couple to the other $\phi_{2}$. The ratio of vacuum
    expectation values (vev) is a free parameter defined by $\tan(\beta)= v_{2}/v_{1}$.
    By requiring that the theory be perturbative we
    obtain a bound on the possible values of $\tan(\beta)$. The Yukawa
    couplings of fourth generation quarks are given by $g_{t'}/\sqrt{2} = m_{t'}/v_{2}$ and $g_{b'}/\sqrt{2} =
    m_{b'}/v_{1}$, where $v^{2} = v^{2}_{1} + v^{2}_{2} = (246
    \textrm{GeV})^{2}$.
    If we approximate $m_{t'} \sim m_{b'} \equiv M \gtrsim 280$ GeV and relate $v_{1}$ and $v_{2}$
    to the Standard Model Higgs vev $v$ in terms of $\tan(\beta)$, then
    the theory will remain perturbative, $g_{t'}^2 < 4\pi$ and $g_{b'}^2 < 4\pi$, only if
    \begin{equation}
        \label{perturbative tan beta requirement}
        \frac{1}{\sqrt{2\pi(v/M)^{2}-1}} < \tan(\beta) < \sqrt{2\pi (v/M)^{2}-1} \ .
    \end{equation}
    Thus for Model II with $M \geq 280$ GeV we
    find $1/2 < \tan(\beta) < 2$. In Model I the $\phi_{1}$
    field does not couple to fermions, which eliminates the upper bound,
    the lower bound however remains unchanged.
    
    Issues of perturbation theory and vacuum stability pose a 
    challenge to four-generation models.  As noted most recently by 
    Kribs \cite{kpst}, in the Standard Model the large Yukawa 
    couplings will cause the scalar self-coupling 
    to either go negative (leading to vacuum instability) or reach a 
    Landau pole well before the GUT scale.  The Yukawa coupling 
    itself can also reach a Landau pole at relatively low scales.  
    Although methods can be found to extend the reach of perturbation 
    theory \cite{murdock}, they do involve addition of new physics 
    just above the TeV scale.  In the two Higgs doublet models, the 
    situation is much more complicated since there are many scalar 
    self-couplings, many other vacua (such as charge-breaking vacua) 
    and many other opportunities for instabilities.  Our approach 
    here is to assume that the two Higgs doublet model is an 
    effective theory below the TeV scale, and will presume that 
    physics above that scale will not substantially affect our results.

    Since the charged Higgs $H^{\pm}$ is a possible decay product of the
    $t'$ through the decay $t' \rightarrow Hb$, and this decay will be
    very difficult to observe, there will be a slight
    suppression in the branching ratio BR($t' \rightarrow Wb$), which is
    shown in Figure 2 as a function of the charged Higgs mass.

    \begin{figure}
        \includegraphics[scale=.75]{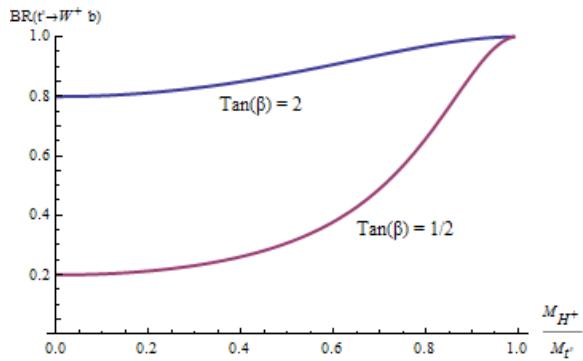}
        \caption{\label{Plot2}Branching Ratio for $t' \rightarrow W b$ as a
            function of the ratio $M_{H^{+}}/M_{t'}$.}
    \end{figure}

    A much more interesting situation arises in Model III.    In Model
    III, there is no discrete symmetry prohibiting tree level flavor
    changing neutral currents.   Initially this appears to be very problematic.
    However if one goes to a basis in which one of the scalar fields gets a
    vacuum expectation value and the other does not, then the couplings
    of the latter, $\phi$, will be, in general, flavor changing. By
    analyzing various mass matrix textures, Cheng and
    Sher \cite{chengsher} argued that fine-tuning can be avoided if the
    flavor changing neutral couplings $\xi_{ij} \overline{f}_{i} f_{j}\phi$
    are given by
    \begin{equation}
        \xi_{ij} = \lambda_{ij}{\sqrt{m_{i}m_{j}}\over v/\sqrt{2}}
    \end{equation}
    and the couplings $\lambda_{ij}$ are of $\mathcal{O}(1)$.   In other words,
    the flavor changing Yukawa couplings are the geometric mean of the
    two Yukawa couplings of the fermions involved.  Model III is 
    defined as the two Higgs doublet model with Yukawa couplings 
    given by the above expression.   Recent studies \cite{recents} of heavy quark 
    mixing and decays have begun exploring interesting regions of 
    parameters space (which depends on the $\lambda_{ij}$ and the 
    relevant Higgs masses).  Note that in this model, 
    the flavor changing couplings of the light quarks are very small,
    and the constraints from kaon physics are not as severe.    Only 
    a few studies of Model III have been done
    \cite{sher99,hou01,hou08}
    involving fourth generation fields.

    In this Model, one would expect an enormous flavor changing coupling
    between the $t'$, the $t$ and the $\phi$.  The coupling, in fact,
    would be substantially larger than the top quark Yukawa coupling.  If
    kinematically accessible, then one would expect $t'\rightarrow t\phi$
    to overwhelmingly dominant $t'$ decays.  Here, $\phi$ can be the
    combination of neutral scalars that is orthogonal to the state that
    gets a vev, or it can be the pseudoscalar.  In either event, this
    decay will dominate.

    The cross section for producing a 400 GeV $t'$ is 15 picobarns
    \cite{mangano}.
    Virtually all of these $t'$s will decay into $t\phi$, leading to a
    dramatic $t\bar{t}\phi\phi$ signature.    If $\phi$ is a pseudoscalar
    or a neutral scalar lighter than about 140 GeV, then it will decay
    into $b\bar{b}$, leading to a $6b, 2W$ final state.   The
    biggest Standard Model background to these events would come from
    double pair production of $t\bar{t}b\bar{b}$.   The cross section for this
    background \cite{bddp} is approximately 2 picobarns, and it gives a
    $4b, 2W$ final state.  If one only looks at events with three or
    more tagged $b$'s and one or more leptons, and assumes a $b$-tag
    efficiency of 40\%, then the signal will pass the cut $20\%$ of the
    time, and the background will pass the cut $8\%$  of the time,
    leading to a signal of 3150 fb and a background of only 160 fb.
    Thus, it appears that the signal will easily be detectable, possibly
    in an early run at the LHC.

    If $\phi$ is a
    neutral scalar heavier than 140 GeV, then it will decay into $WW$
    leading to a $2b, 6W$ final state.  Here, if one looks at events with
    three or more leptons and one or more tagged $b$'s, then $12\% $
    of the decays will pass the cuts, leading to an event rate of
    1.8 picobarns.  We know of no standard model background that comes close
    to this signal.  This signal would also be easily detectable at the
    LHC.

\section{Conclusions}

    There continues to be interest in the phenomenology of a sequential
    fourth generation.  Yet almost all discussion have been in the
    context of the Standard Model. 
    In this paper, we have explored the phenomenology of the fourth
    generation, focusing on the $t'$ quark, in extensions
    of the Standard Model.  In the
    Randall-Sundrum model, the decay $t'\rightarrow tZ$ can occur at a
    rate approaching one percent, which should be detectable at the LHC.
    In two-Higgs doublet model, one gets a suppression of the branching
    ratio $t'\rightarrow Wb$ in Models I and II, but in Model III the
    decay of $t'\rightarrow t\phi$, where $\phi$ is a scalar or
    pseudoscalar, dominates the decay, leading to spectacular
    $t\bar{t}\phi\phi$ signatures at very large rates, which could be
    detected during the early months of running at the LHC.

\begin{acknowledgments}
    We are grateful to Chris Carone, Graham Kribs and Xerxes Tata for
    useful discussions.  This work was supported by the National Science
    Foundation grant NSF-PHY-0757481.
\end{acknowledgments}

\end{document}